\documentclass[conference]{IEEEtran}
\IEEEoverridecommandlockouts
\usepackage{booktabs}
\usepackage{cite}
\usepackage{amsmath,amssymb,amsfonts}
\usepackage{algorithmic}
\usepackage{graphicx}
\usepackage{textcomp}
\usepackage{xcolor}
\usepackage{subfig}
\def\BibTeX{{\rm B\kern-.05em{\sc i\kern-.025em b}\kern-.08em
    T\kern-.1667em\lower.7ex\hbox{E}\kern-.125emX}}

\usepackage{bm}
\newtheorem{theorem}{Theorem}

\newtheorem{definition}{Definition}

\usepackage{amsmath}
\DeclareMathOperator*{\argmax}{arg\,max}

\DeclareRobustCommand*{\IEEEauthorrefmark}[1]{%
  \raisebox{0pt}[0pt][0pt]{\textsuperscript{\footnotesize #1}}%
}

\begin{document}

\title{
    Social Welfare Maximization for Collaborative Edge Computing: A Deep Reinforcement Learning-Based Approach
}

\author{\IEEEauthorblockN{Xingqiu He\IEEEauthorrefmark{1},
        Yuhang Shen\IEEEauthorrefmark{2},
        Hongxi Zhu\IEEEauthorrefmark{2}, 
        Sheng Wang\IEEEauthorrefmark{2},
        Chaoqun You\IEEEauthorrefmark{1} and Tony Q.S. Quek\IEEEauthorrefmark{1}}
        \IEEEauthorblockA{
            \IEEEauthorrefmark{1}
            Information Systems Technology and Design, Singapore University of Technology and Design  \\
            \IEEEauthorrefmark{2}
            School of Information and Communication Engineering, University of Electronic Science and Technology of China\\
}
}

\maketitle

\begin{abstract}
    Collaborative Edge Computing (CEC) is an effective method that 
    improves the performance of Mobile Edge Computing (MEC) systems by offloading computation tasks from busy edge servers (ESs) to idle ones.
    However, ESs usually belong to different MEC service providers so they have no incentive to help others.
    To motivate cooperation among them, 
    this paper proposes a cooperative mechanism where idle ESs can earn extra profits by sharing their spare computational resources.
    To achieve the optimal resource allocation,
    we formulate the social welfare maximization problem as a Markov Decision Process (MDP) and decompose it into two stages
    involving the allocation and execution of offloaded tasks.
    The first stage is solved by extending the well-known Deep Deterministic Policy Gradient (DDPG) algorithm.
    For the second stage, we first show that we only need to decide the processing order of tasks and the utilized computational resources.
    After that, we propose a dynamic programming and a Deep Reinforcement Learning (DRL)-based algorithm
    to solve the two types of decisions, respectively.
    Numerical results indicate that our algorithm significantly improves social welfare under various situations.

\end{abstract}


\section{Introduction}
Recently, the proliferation of mobile devices and the commercialization of 5G technology
have driven the development of many mobile applications, such as virtual reality and mobile gaming.
However, restricted by the physical size and manufacturing cost, existing mobile devices are usually unable to 
provide sufficient computational power for these computation-intensive applications.
To address this issue, Mobile Edge Computing (MEC) \cite{hu2015mobile} has been proposed as a new computing paradigm that 
enables mobile devices to offload their computation tasks to nearby edge servers (ESs).
Extensive research has shown that MEC substantially reduces the computation delay of tasks
and thus improves the user experience.

A major problem in MEC is the computational resources in each ES
are very limited. Hence, ESs often cannot process all incoming computation tasks timely 
when facing bursty task arrivals.
Fortunately, this problem can be significantly alleviated via the 
Collaborative Edge Computing (CEC) framework \cite{tran2017collaborative, shi2016edge, wang2018enabling},
which balances the workload among ESs by offloading computation tasks from busy ESs to idle ones.
However, since ESs usually belong to different MEC service providers, they have no incentive to help others.
To encourage cooperation among them, the authors in \cite{he2021shapley} and \cite{hou2021incentive} 
have proposed two cooperative mechanisms for CEC networks.
A deficiency of the two mechanisms is they both
assume that ESs are willing to contribute all of their computational resources
and be managed by a central controller.
In practice, however, self-interested ESs only help others if there are extra computational resources after processing
their own tasks.
Besides, ESs are reluctant to hand over their control to others due to security considerations.

To address the above issue, this paper proposes a more practical cooperative mechanism that 
enables ESs to retain control over themselves.
Instead of managing all ESs with a central controller,
our mechanism lets ESs decide the amount and price of computational resources they are willing to share.
When an ES is overloaded and desires to offload a computation task, it posts an offloading request that contains
the task's information and the amount of reward it is willing to pay.
Based on the information of shared resources and offloading requests,
we can compute the optimal scheduling scheme that maximizes social welfare.
The main contributions of our work are summarized as follows:
\begin{itemize}
    \item We propose a cooperative mechanism for CEC networks that enables ESs to retain control over themselves.
        To achieve the optimal resource allocation, we formulate the social welfare maximization problem as a Markov Decision Process (MDP).
    \item We decompose the original MDP into two stages involving task allocation and task execution.
        The first stage is solved by extending the Deep Deterministic Policy Gradient (DDPG) algorithm.
        The second stage is further decomposed into two procedures and then solved by dynamic programming and
        Deep Reinforcement Learning (DRL), respectively.
    \item We conduct extensive numerical simulations to validate the effectiveness of the proposed algorithm.
        Our algorithm is compared with two benchmarks and numerical results show that our algorithm
        significantly improves the system performance under various settings.
\end{itemize}

\section{System Model and Problem Formulation} \label{section:system_model}
In this section, we first describe our system model and the cooperative mechanism among ESs.
After that, we formulate the social welfare maximization problem as an MDP.

\subsection{System Model and Cooperative Mechanism}
\begin{figure}[!t]
    \centering
    \includegraphics[width=0.45\textwidth]{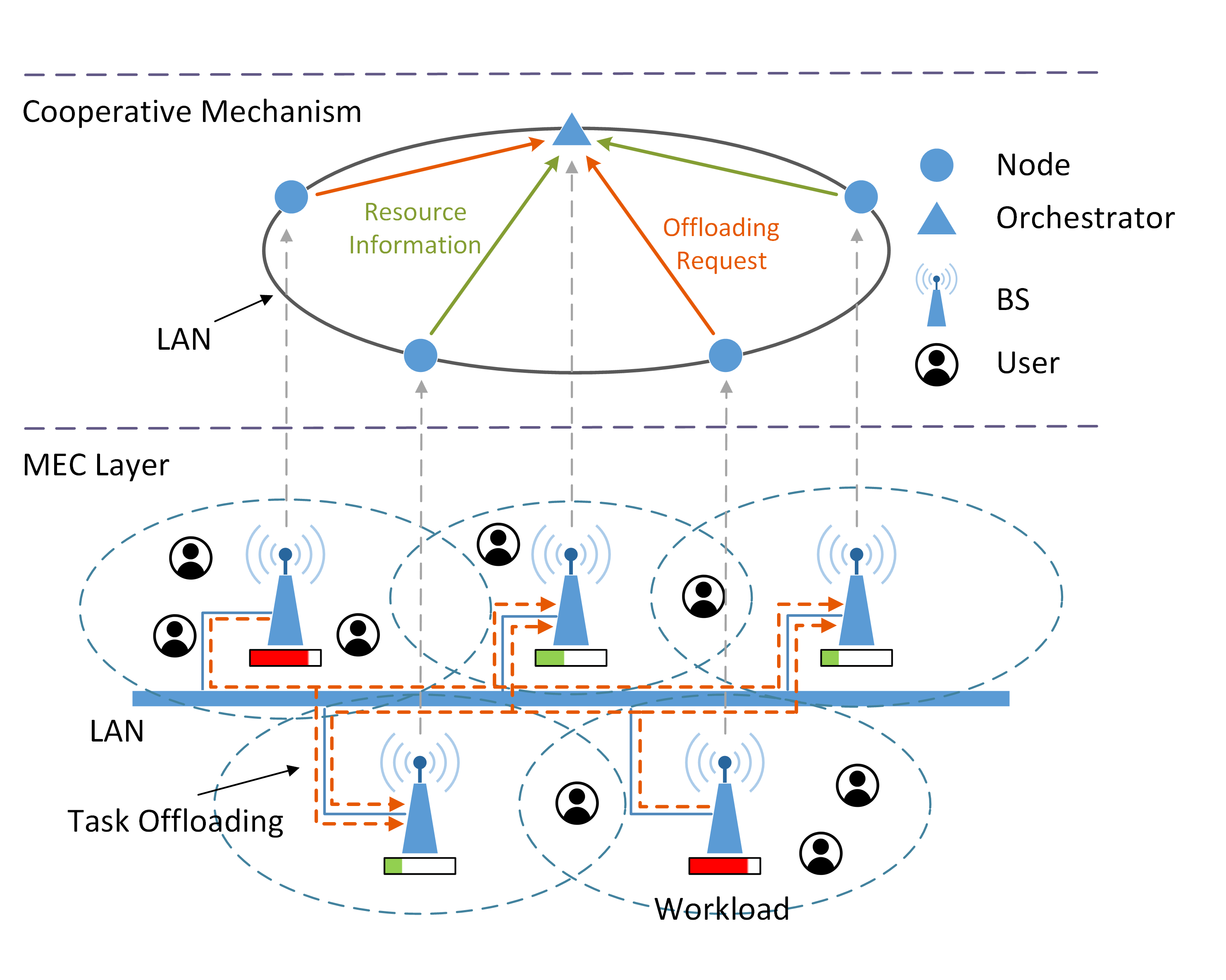}
    \caption{A simple example of the considered system.}
    \label{fig:system}
\end{figure}
As shown in Fig.\ref{fig:system}, we consider a CEC network with $N$ base stations (BSs) indexed by $\mathcal{N} = \{ 1, \dots, N \}$.
Each BS is endowed with computational capabilities and serves as an ES.
To provide a better user experience, overloaded BSs may wish to offload part of their tasks to other BSs.
Meanwhile, idle BSs are willing to share their computational resources for extra profits.
Therefore, offloading tasks from busy BSs to idle ones results in a win-win situation.
To maximize the benefits of cooperation, we design a orchestrator to collect necessary information
(e.g. the amount and price of spare computational resources) and compute the optimal task scheduling scheme.
The obtained task scheduling scheme is distributed to the corresponding BSs so that they can cooperate in
a peer-to-peer manner.
Notice that the orchestrator is different from the central controller in \cite{he2021shapley} and \cite{hou2021incentive},
as it does not control the behavior of BSs.
In practice, the orchestrator can be implemented by any entity with sufficient communication and computational capabilities
and is usually elected by all BSs or designated by the governmental institution.

For convenience, the time horizon is divided into multiple slots of equal length $\Delta t$.
In each time slot $t$, the cooperation among BSs is conducted through the following steps.

\subsubsection{Revealing Spare Resources}
At the beginning of slot $t$, idle BSs should report the available amount and unit price of their spare computational
resources\footnote{We only consider CPU frequency in this paper, but our mechanism can be easily extended to include other types of computational resources.} 
over the next $\delta$ slots.
We assume the amount and price of the spare resources remain constant within each slot, but may be different across
multiple slots.
Therefore, the resource information declared by BS $i$ at slot $t$ can be represented by
two $\delta$-demension vectors $C^{t}_i$ and $p^{t}_i$,
where $C^t_i(t')$ and $p^t_i(t')$ denote the available CPU frequency and corresponding unit price 
at slot $t'\in [t, t+\delta-1]$.
Notice that $C_i^{t}(t')$ could be zero if BS $i$ is busy at that slot.

\subsubsection{Posting Offloading Request}
When BS $i$ is overloaded and wishes to offload task $\tau$ to its neighbors, it should post an offloading request to the orchestrator.
The request contains the task's workload $w_{\tau}$ and utility function $u_{\tau}(l)$,
where $w_{\tau}$ is the number of CPU cycles required to complete $\tau$
and $u_{\tau}(l)$ returns the utility value of $\tau$ when its latency is $l$. 
In this paper, we assume the utility function is linear, i.e. $u_{\tau}(l) = u^0_{\tau} - \alpha_{\tau}l$,
where $u^0_{\tau}$ is the maximum utility and $\alpha_{\tau}$ is the punishment coefficient associated with latency $l$.

\subsubsection{Computing Scheduling Scheme}
After collecting the necessary information in the previous two steps,
the orchestrator computes scheduling schemes for the received offloading requests.
For task $\tau$ whose request is posted at slot $t$, its scheduling scheme $s_{\tau} = (i_{\tau}, f_{\tau})$ consists of
two components, where $i_{\tau}$ is the target BS that $\tau$ will be offloaded to and
$f_{\tau}$ is a $\delta$-demension vector specifying the CPU frequency allocated to task $\tau$ in the next $\delta$ slots 
(i.e. time interval $[t, t+\delta-1]$).
In the rest of this paper, $i_{\tau}$ and $f_{\tau}$ are referred to as the task allocation and task execution decisions, respectively.
If the utility of $\tau$ is relatively low, we can also reject its offloading request.
For convenience, we represent the rejection of the request by setting $i_{\tau} = 0$.
Based on $f_{\tau}$, we can calculate task $\tau$'s ending time $t^e_{\tau} = \min\{t':\sum_{\hat{t} \leq t'} f_{\tau}(\hat{t})\Delta t \geq w_{\tau} \}$
and latency $l_{\tau} = t^e_{\tau} - t$.
We define the surplus of task $\tau$ as its utility minus the execution cost, i.e. 
$r_{\tau} = u_{\tau}(l_{\tau}) - \sum_{t'=t}^{t+\delta-1} p^t_{i_{\tau}}(t') f_{\tau}(t')$.
The orchestrator's objective is to find the optimal scheduling scheme that maximizes social welfare,
which is defined as the long-term surplus of all offloaded tasks.

\subsection{Problem Formulation}
In practical situations, the scheduling scheme in the current time slot will influence the BSs' behaviors in the future.
For example, if we offload more tasks to BS $i$, then it is more likely to have fewer spare computational resources
in the next few time slots.
The relationship between current scheduling decisions and subsequent behaviors of BSs can be captured by the state transition process in MDP.
Therefore, we can model the social welfare maximization problem as an MDP, 
which is defined by a $4$-tuple $<\mathcal{S}, \mathcal{A}, P, R>$, where
\begin{itemize}
    \item $\mathcal{S}$ represents the state space.
        Let $S(t)\in\mathcal{S}$ be the system state at time slot $t$.
        In our model, $S(t)$ contains the related information collected from BSs,
        i.e. $S(t) = \{ (C^t_i, p^t_i)_{i\in \mathcal{N}}, (w_{\tau}, u_{\tau}(l))_{\tau \in \mathcal{T}(t)} \}$,
        where $\mathcal{T}(t)$ is the set of all offloading requests at slot $t$.
    \item $\mathcal{A}$ is the action space.
        The action at slot $t$, denoted by $A(t)$, specifies the scheduling of each task in $\mathcal{T}(t)$,
        i.e. $A(t) = (s_{\tau})_{\tau \in \mathcal{T}(t)}$.
        Notice that in each BS the CPU frequency allocated to tasks should not exceed the available amount.
        Hence, $A(t)$ should satisfy
        \begin{equation*}
            \sum_{\tau:i_{\tau} = i} f_{\tau}(t') \leq C^t_i(t'), \quad \forall i\in\mathcal{N}, \forall t'\in [t, t+\delta-1].
        \end{equation*}
    \item $P(S(t+1) | S(t), A(t))$ is the probability that the system state switches to $S(t+1)$ if we take action $A(t)$ in state $S(t)$.
        For self-interested BSs, they are willing to help others only if there are extra computational resources after fulfilling their own business.
        Therefore, the next state $S(t+1)$ depends on each BS's task arrival process and pricing mechanism, which is unknown to the orchestrator.
        As a result, the transition probability is also unknown in advance and must be learned from experience.
    \item $R(t) = R(S(t), A(t))$ gives the immediate reward for being in state $S(t)$ and taking action $A(t)$.
        In our problem, $R(t)$ 
        is the total surplus of tasks in $\mathcal{T}(t)$, defined as
        $R(t) = \sum_{\tau \in \mathcal{T}(t)} r_{\tau}$.
        Our objective is to maximize the long-term social welfare, which is the
        sum of discounted future reward, defined as $G = \lim_{T\to\infty} \sum^{T}_{t=0} \gamma^t R(t)$,
        where $\gamma \in [0,1]$ is the discount factor.
\end{itemize}

Since the transition probability is unknown, the formulated MDP cannot be solved directly and 
must resort to reinforcement learning (RL) algorithms.
However, due to the continuous state space and action space, our problem cannot be solved by the 
conventional tabular-based RL algorithms \cite{auer2008near}.
To overcome this problem, we will first decompose the formulated MDP into two stages
and then use Deep Neural Network (DNN)-based RL algorithms to solve it.

\section{Deep Reinforcement Learning-Based Scheduling Algorithm} \label{section:algorithm}
In this section, we design a DRL-based algorithm to solve the formulated MDP.
Our algorithm consists of two stages.
The first stage decides the target BS $i_{\tau}$ for each task $\tau \in \mathcal{T}(t)$.
The second stage calculates the optimal task execution $f_{\tau}$ based on the decisions made in the first stage.

\subsection{First Stage: Task Allocation} \label{subsection:first_stage}
The first stage only decides the allocation of tasks in $\mathcal{T}(t)$
so the action in the first stage $A^a(t)$ can be defined as $A^a(t) = (i_{\tau})_{\tau \in \mathcal{T}(t)}$.
The action space of $A^a(t)$ is denoted by $\mathcal{A}^a$.
Since the task execution $A^e(t) = (f_{\tau})_{\tau\in\mathcal{T}(t)}$ calculated in the second stage relies on $A^a(t)$,
we can regard $A^e(t)$ as a function of $A^a(t)$, denoted by $A^e(t) = g(A^a(t))$.
Recall that $A(t) = (s_{\tau})_{\tau\in\mathcal{T}(t)} = (A^a(t), A^e(t))$,
so we can define the reward of $A^a(t)$ as the reward of $A(t) = (A^a(t), g(A^a(t)))$.
Hence, the reward function of the first stage can be defined as
$R^a(t) = R^a(S(t), A^a(t)) = R(S(t), (A^a(t), g(A^a(t))))$.
Then the MDP problem in the first stage is represented by the $4$-tuple $<\mathcal{S}, \mathcal{A}^a, P, R^a>$.
However, this problem is still hard to solve due to the exponentially large state and action spaces.

\begin{figure}[!t]
    \centering
    \includegraphics[width=0.49\textwidth]{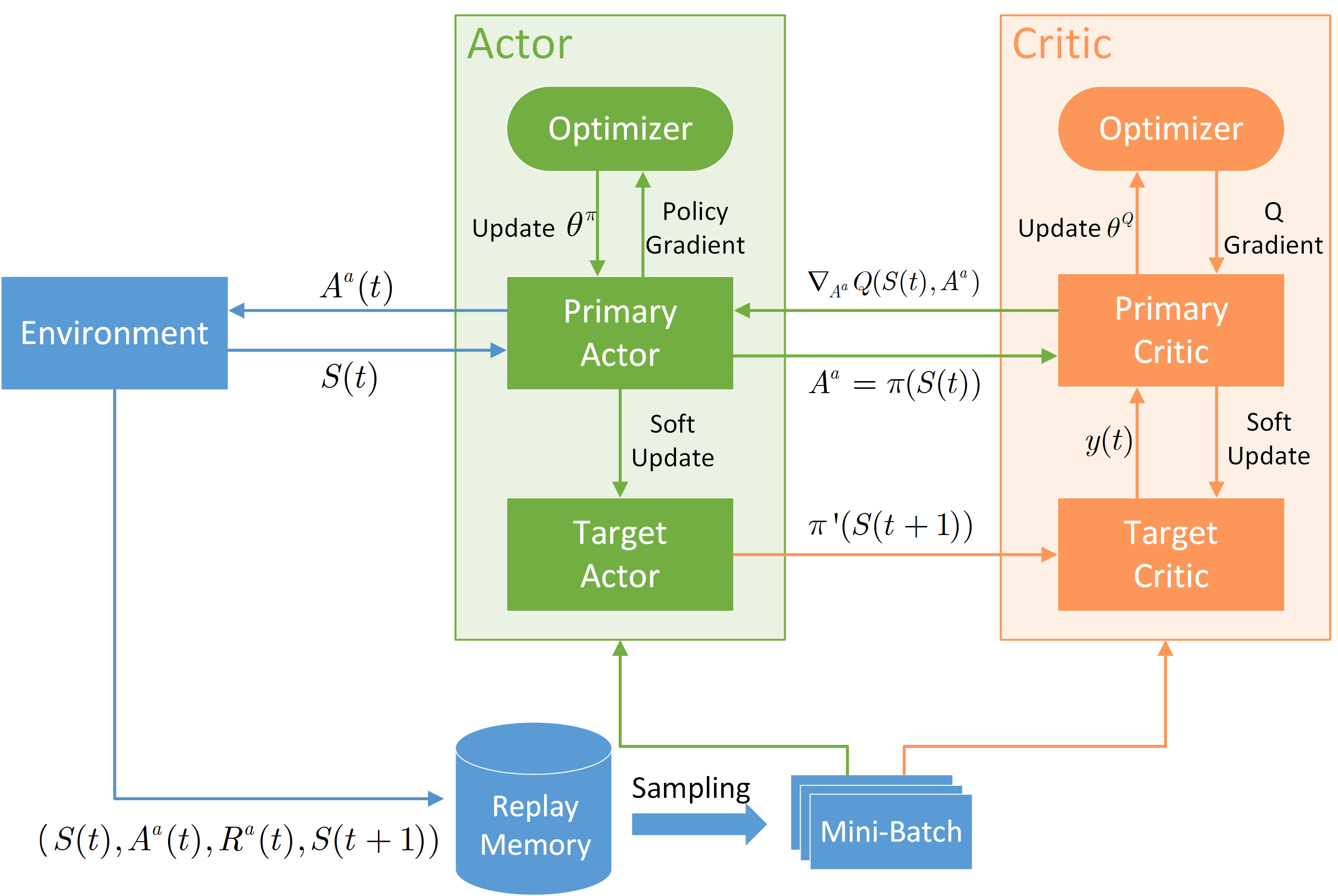}
    \caption{Structure of the DRL algorithm in the first stage.}
    \label{fig:ddpg}
\end{figure}
The recent success in applying DNNs to RL algorithms has driven the development of many DRL methods.
In this paper, we extend the Deep Deterministic Policy Gradient (DDPG) algorithm \cite{ddpg}, which is capable to handle 
continuous high-dimensional state and action spaces, to solve the task allocation problem.
The structure of our algorithm is illustrated in Fig.\ref{fig:ddpg}.
Let $\pi: \mathcal{S} \to \mathcal{A}^a$ be any deterministic policy that returns a feasible task allocation for any given state $S$.
Then we can define the corresponding action-value function as
\begin{align*}
    Q^{\pi}&(S(t), A^a(t))  \\
    &= \mathbb{E} \left[ \lim_{T\to\infty} \sum_{i=t}^{T} \gamma^{i-t} R^a(S(i), \pi(S(i))) \,|\, S(t), A^a(t) \right]
\end{align*}
where the expectation is taken over the state transition probability.
Clearly, $Q^{\pi}(S(t), A^a(t))$ is the long-term return after taking action $A^a(t)$ in state $S(t)$ and thereafter following policy $\pi$.
According to the Bellman optimality equation, the action-value function under the optimal policy $\pi^*$ satisfies
\begin{align}
    Q&^{\pi^*}(S(t), A^a(t)) \notag \\
    &= \mathbb{E} \left[ R^a(t) + \gamma \max_{A^a(t+1)} Q^{\pi^*}\left(S(t+1), A^a(t+1)\right) \right] \label{eq:bellman}
\end{align}
and the optimal policy can be computed by
\begin{align*}
    \pi^*(S(t)) = \argmax_{A^a(t)} Q^{\pi^*}(S(t), A^a(t)).
\end{align*}
Unfortunately, under high-dimensional state and action spaces, it is computationally impossible to directly 
learn the optimal action-value function and calculate the optimal policy
using conventional RL algorithms like Q-learning \cite{watkins1992q}.

To avoid the curse of dimensionality, DDPG utilizes two DNNs to approximate the policy function and the action-value function.
The two DNNs are usually called the primary actor network (whose parameter is denoted by $\theta^{\pi}$)
and the primary critic network (whose parameter is denoted by $\theta^{Q}$).
Then the two approximated functions can be represented by $\pi(S \,|\, \theta^{\pi})$ and $Q(S, A^a \,|\, \theta^{Q})$, respectively.
Based on the Bellman optimality equation \eqref{eq:bellman}, we can define the loss of the primary critic network as
\begin{equation*}
    L(\theta^Q) = \mathbb{E} \left[ \left( Q(S(t), A^a(t) \,|\, \theta^Q) - y(t) \right)^2 \right]
\end{equation*}
where
\begin{equation}
    y(t) = R^a(t) + \gamma Q(S(t+1),\pi(S(t+1)) \,|\, \theta^Q). \label{eq:y}
\end{equation}
Notice that when $L(\theta^Q) = 0$, the corresponding $Q(S, A^a|\theta^Q)$ satisfies the Bellman optimality equation \eqref{eq:bellman}
and thus equals $Q^{\pi^*}$.
Therefore, to approximate $Q^{\pi^*}$ we only need to choose the parameter $\theta^Q$ that minimizes $L(\theta^Q)$.
On the other hand, the result in \cite{silver2014deterministic} shows 
the update of the parameter $\theta^{\pi}$ relies on the sampled policy gradient and can be computed by
\begin{equation*}
    \nabla_{\theta^{\pi}}J \approx \mathbb{E} \left[ \nabla_{A^a} Q(S(t), A^a | \theta^Q) |_{A^a=\pi(S(t))} \nabla_{\theta^{\pi}} \pi(S(t) | \theta^{\pi}) \right].
\end{equation*}

To improve the algorithm's performance, DDPG applies two extra techniques during the training processes of DNNs.
First, in order to guarantee the training data are independently and identically distributed, DDPG stores historical experience in a replay buffer
and uniformly samples a minibatch at each training step.
Second, when minimizing the critic loss $L(\theta^Q)$, we try to make the action-value function as close to $y(t)$ as possible.
However, according to \eqref{eq:y}, $y(t)$ depends on the same parameter $\theta^Q$ we are trying to optimize
and this may cause a stability problem during the update.
To address this issue, DDPG creates a copy of the primary actor and critic network, denoted by $\pi'(S \,|\, \theta^{\pi'})$ and $Q'(S, A^a \,|\, \theta^{Q'})$.
The purpose of $\pi'$ and $Q'$ is to calculate the target $y(t)$ so they are called the target actor and critic network.
The parameters of $\pi'$ and $Q'$ are updated to slowly track the primary networks.
Their update formula is given by
\begin{align*}
    \theta^{Q'} &= \omega \theta^{Q} + (1-\omega) \theta^{Q'} \\
    \theta^{\pi'} &= \omega \theta^{\pi} + (1-\omega) \theta^{\pi'}.
\end{align*}
With the help of target networks, $y(t)$ can be computed as
\begin{equation*}
    y(t) = R^a(t) + \gamma Q'(S(t+1),\pi'(S(t+1)|\theta^{\pi'}) \,|\, \theta^{Q'}),
\end{equation*}
which no longer depends on $\theta^Q$ and thus avoids the stability problem.

The DDPG algorithm is originally designed for MDP problems with fixed state dimensions.
In our problem, however, the length of $S(t)$ is variable because the number of offloading requests changes across different time slots.
To solve this issue, we divide the tasks in $\mathcal{T}(t)$ into $\lceil |\mathcal{T}(t)| / K \rceil$ groups 
(where $K$ is a tunable parameter) and each group contains $K$ tasks.
If the last group has insufficient tasks, we fill it with dummy tasks whose workload and utility are both zero.
Then the DDPG is applied repeatedly to decide the allocation of each group of tasks.
Since the allocation decisions are categorical data, we utilize the one-hot encoding in the output layer.
That is, our output layer has $K(N+1)$ components and component $i(N+1)+j$ represents allocating task $i$ to BS $j \in \{0\}\cup \mathcal{N}$.
Since DDPG returns continuous values, we apply the softmax function to the $N+1$ outputs associated with each task to obtain a feasible allocation.

\subsection{Second Stage: Task Execution} \label{subsection:second_stage}
After obtaining the task allocation decision $A^a(t)$, we still need to compute the task execution strategy $A^e(t)$ to produce a feasible scheduling.
Let $\mathcal{T}_i(t) = \{ \tau \in \mathcal{T}(t): i_{\tau} = i \}$ be the set of tasks allocated to BS $i$,
then the task execution problem on BS $i$ can be formulated as
\begin{align}
    \max_{f_{\tau}}\quad &\sum_{\tau \in \mathcal{T}_i(t)} r_{\tau} = \sum_{\tau \in \mathcal{T}_i(t)} \left[ u_{\tau}(l_{\tau}) - \sum_{t'=t}^{t+\delta-1} p^t_i(t')f_{\tau}(t') \right] \label{tep} \\
    s.t.\quad &\sum_{\tau\in\mathcal{T}_i(t)} f_{\tau}(t') \leq C^t_i(t'), \quad \forall t'\in [t, t+\delta-1] \tag{\ref{tep}{a} \label{tep:capacity}} \\
    &f_{\tau}(t') \geq 0, \quad \forall t'\in [t, t+\delta-1] \tag{\ref{tep}{b} \label{tep:var}}
\end{align}
Let $A^e_i(t) = (f_{\tau})_{\tau\in\mathcal{T}_i(t)}$ be the task execution decisions associated with BS $i$.
The following two theorems characterize the optimal solutions of problem \eqref{tep}.
\begin{theorem}
    There exists an optimal solution $A^{e,*}_i(t)$ 
    such that no tasks are processed simultaneously, i.e. a new task will be processed only if the current task is completed.
    \label{theorem:no_simul}
\end{theorem}

\begin{figure}[t]
\centering
\subfloat[Simultaneous]{\includegraphics[width=0.3\textwidth]{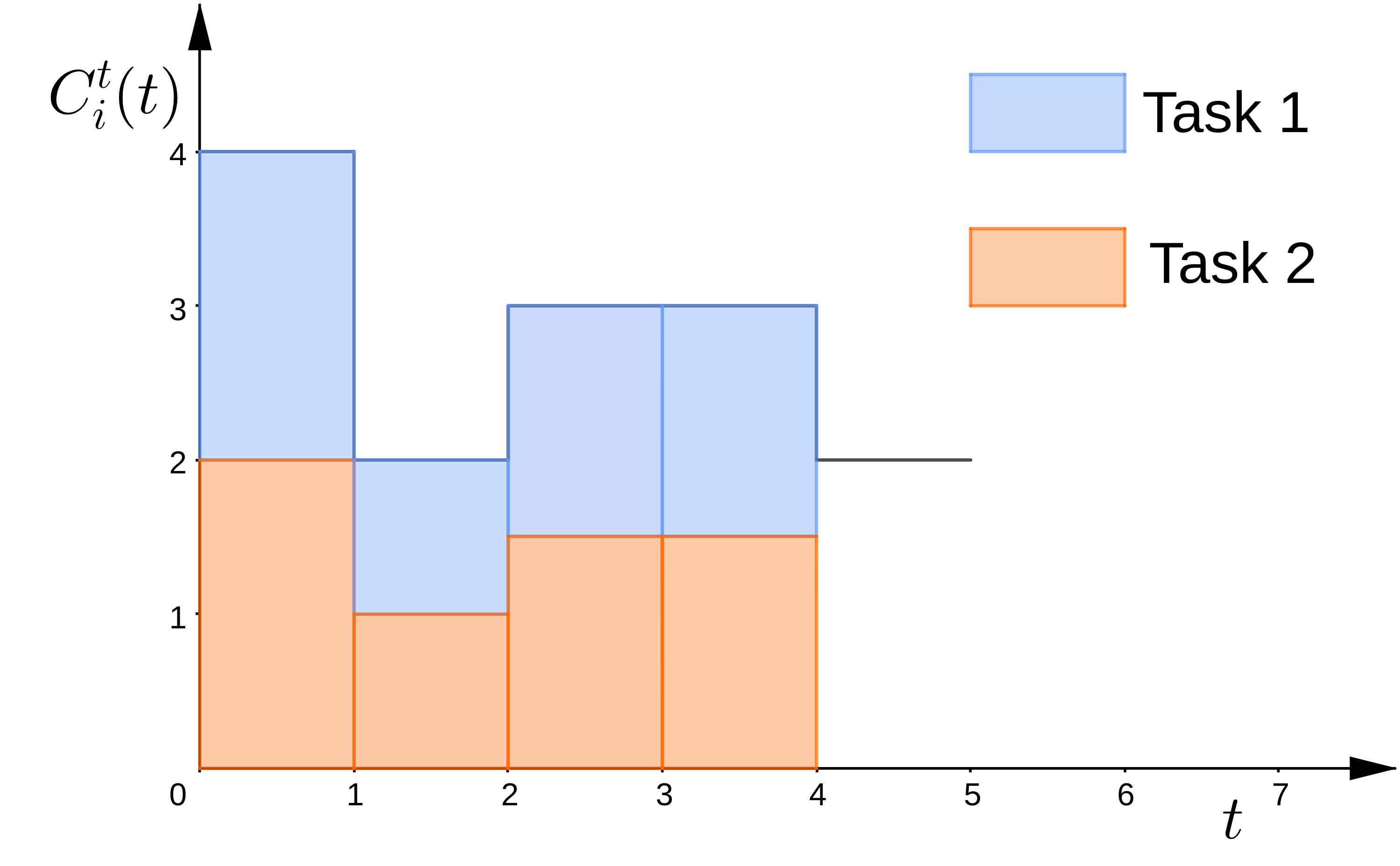} \label{fig:theorem1_1}}
\vfil
\subfloat[Sequential]{\includegraphics[width=0.3\textwidth]{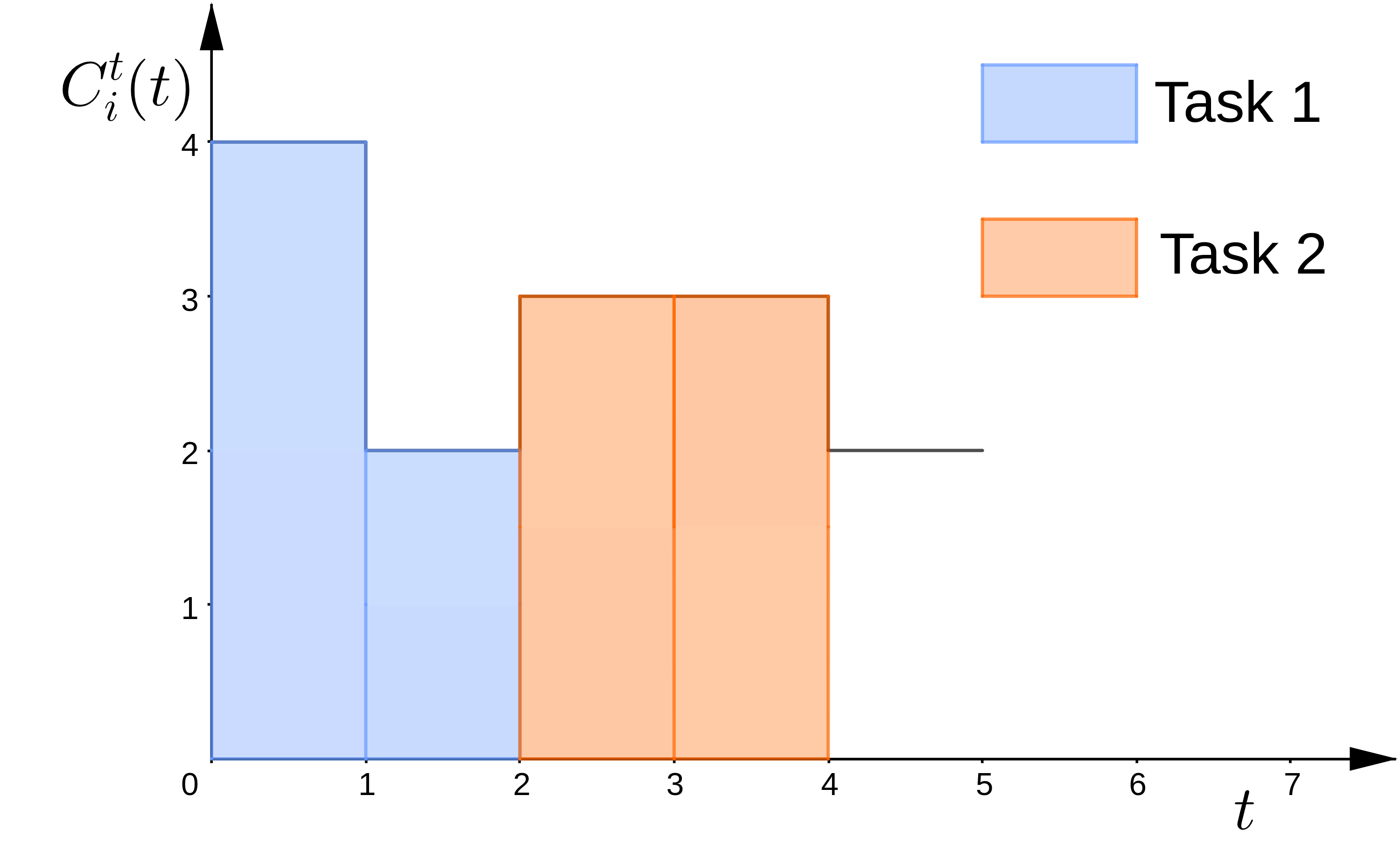} \label{fig:theorem1_2}}
\caption{An illustration of the two processing strategies.}
\label{fig:theorem1}
\end{figure}

The intuition behind Theorem \ref{theorem:no_simul} can be illustrated by a simple example.
As shown in Fig.\ref{fig:theorem1}, we consider two tasks with an equal workload.
If processed simultaneously, both tasks will be completed in $4$ slots.
However, if we handle them separately, then the latency of the first task reduces to $2$ slots.
By devoting all computation power to one task at a time, the processing time of each task is minimized
and thus shortens the latency of tasks at the front of the queue.
Theorem \ref{theorem:no_simul} is the natural generalization of this result to the multi-task cases and the detailed proof is omitted for brevity.

Before introducing the second theorem, we first define a special class of task execution decisions.
\begin{definition}
    A task execution decision $A^e_i(t)$ is called binary if for any slot $t'\in[t, t+\delta-1]$, we have either $\sum_{\tau\in\mathcal{T}_i(t)} f_{\tau}(t') = 0$
    or $\sum_{\tau\in\mathcal{T}_i(t)} f_{\tau}(t') = C^t_i(t')$.
\end{definition}

According to the definition, if a binary task execution decision $A^e_i(t)$ processes tasks at slot $t'$, 
then it will use up all available computational power in that slot.
The following theorem states that binary task execution decisions
can achieve the best objective value if the duration of each slot $\Delta t$ approaches zero.
\begin{theorem}
    Let $R^e(A^e_i(t))$ be the objective value of problem \eqref{tep} under decision $A^e_i(t)$.
    Suppose $\hat{A}^{e,*}_i(t)$ is the optimal \emph{binary} task execution decision
    and ${A}^{e,*}_i(t)$ is the optimal task execution decision (not necessarily binary).
    Then we have $\lim_{\Delta t\to 0} R^e(\hat{A}^{e,*}_i(t)) = R^e(A^{e,*}_i(t))$.
    \label{theorem:useup}
\end{theorem}
\begin{IEEEproof}
    Please see Appendix \ref{appendix:proof_useup}.
\end{IEEEproof}

In MEC systems, in order to satisfy the stringent quality of service (QoS) requirements of real-time applications \cite{hu2015mobile},
the duration of each slot is usually very short (e.g. several milliseconds).
Hence, according to Theorem \ref{theorem:useup}, we only need to focus on binary task execution decisions.

Based on the result in Theorem \ref{theorem:no_simul}, one can easily verify that a binary task execution decision $A^e_i(t)$ 
is uniquely determined once we have specified
the processing order of tasks (denoted by $\sigma_i(t)$) and which time slots are utilized (denoted by $\bm{x}_i(t)$, where the $j$-th component
$x_i^j(t)\in\{0,1\}$ is a binary variable indicating whether slot $t+j-1$ is utilized).
Therefore, to obtain the optimal $A^{e,*}_i(t)$, we only need to find the corresponding $\sigma^*_i(t)$ and $\bm{x}^*_i(t)$.
Unfortunately, this problem is strongly NP-hard even if (i) the capacity of each slot $C^t_i(t')$ is equal
and (ii) there are only two different prices for the computational resources \cite{wang2005preemptive}.
To solve this problem, in the rest of this subsection, we propose a DRL-based approximate algorithm
that produces close-to-optimal solutions efficiently.

\subsubsection{Slot Utilization}
We first show that for an arbitrary processing order $\sigma_i(t)$, we can calculate the corresponding optimal $\bm{x}^*_i(t)$
in polynomial time by dynamic programming.
Let $Z(j,t_1)$ be the optimal total surplus of the first $j$ tasks in $\sigma_i(t)$
when the $j$-th task is completed at slot $t_1$.
Then the Bellman equation of $Z(j, t_1)$ can be expressed as
\begin{align}
    Z(j,t_1) = \max_{t<t_0<t_1} Z(j-1, t_0) + z(j, [t_0+1, t_1])
    \label{eq:bellman2}
\end{align}
where $z(j, [t_0+1, t_1])$ denotes the maximum surplus of the $j$-th task when it is processed after $t_0$ and finished at slot $t_1$.
Since the finishing time of the $j$-th task is fixed, the utility of the $j$-th task is also fixed.
Therefore, maximizing the surplus is equivalent to minimizing the execution cost.
This can be realized by 
sorting $p^t_i(t')$ for all $t'\in[t_0+1,t_1]$ in ascending order and then
choosing time slots with the lowest prices until their total computational power exceeds the $j$-th task's workload.
Therefore, the time complexity for calculating $z(j, [t_0+1, t_1])$ equals the time complexity of the sorting algorithm, 
which is $O((t_1-t_0)\log(t_1-t_0)) = O(\delta \log \delta)$.
Since $1 \leq j \leq |\mathcal{T}_i(t)|$ and $t_1 \in [t, t+\delta-1]$, one can further derive that the time complexity 
for calculating all possible $Z(j,t_1)$ is $O(|\mathcal{T}_i(t)|\delta^3\log \delta)$.
According to the definition of $Z(j,t_1)$,
the optimal total surplus for completing all tasks in $\mathcal{T}_i(t)$ is $\max_{t_1\in[t,t+\delta-1]} Z(|\mathcal{T}_i(t)|, t_1)$.
The corresponding optimal slot utilization $\bm{x}^*_i(t)$ can be immediately obtained by examining
the Bellman equation \eqref{eq:bellman2}.

It should be noted that the equation \eqref{eq:bellman2} is not the actual Bellman equation because there may exist unused computational resources
in slot $t_0$ and if these resources are used to process the $j$-th task, we may obtain a higher surplus.
However, as explained in the proof of Theorem \ref{theorem:useup}, the extra surplus brought by a single slot is negligible when $\Delta t$ is small.
Hence, the $Z(j,t_1)$ calculated by \eqref{eq:bellman2} is very close to the real value so we can regard \eqref{eq:bellman2} as a proper approximation of the actual Bellman equation.

\subsubsection{Processing Order}
After obtaining the optimal $\bm{x}^*(t)$, the next natural question is which $\sigma_i(t)$ should we use?
For convenience, we define the total execution cost $\Gamma_e(A^e_i(t))$ as the sum of the latency cost
$\Gamma_l(A^e_i(t)) = \sum_{\tau\in\mathcal{T}_i(t)} \alpha_{\tau} l_{\tau}$ and the utilization cost
$\Gamma_u(A^e_i(t)) = \sum_{\tau\in\mathcal{T}_i(t)} \sum_{t'=t}^{t+\delta-1} p^t_i(t')f_{\tau}(t')$.
Clearly, we have $R^e(A^e_i(t)) = \sum_{\tau\in\mathcal{T}_i(t)} u^0_{\tau} - \Gamma_e(A^e_i(t))$ so maximizing $R^e$
is equivalent to minimizing $\Gamma_e$.
According to the results in \cite{hohn2015performance, epstein2012universal, chen2021optimal},
given a tunable parameter $\epsilon>0$, we can construct a processing order $\sigma_i'(t)$ in polynomial time so that 
$\Gamma_e(\sigma_i'(t), \bm{x}_i(t)) \leq (4+\epsilon) \Gamma_e(\sigma_i(t), \bm{x}_i(t))$ for any
processing order $\sigma_i(t)$ and slot utilization $\bm{x}_i(t)$.
Let $\bm{x}_i'(t)$ be the optimal slot utilization of $\sigma_i'(t)$ computed by the dynamic programming described above
and $(\sigma_i^*(t), \bm{x}_i^*(t))$ be the optimal execution decision.
Since $\bm{x}_i'(t)$ is optimal with respect to $\sigma_i'(t)$,
we have $\Gamma_e(\sigma_i'(t), \bm{x}_i'(t)) \leq \Gamma_e(\sigma_i'(t), \bm{x}_i^*(t))$.
Moreover, due to the property of $\sigma_i'(t)$, we also have
$\Gamma_e(\sigma_i'(t), \bm{x}_i^*(t)) \leq (4+\epsilon) \Gamma_e(\sigma_i^*(t), \bm{x}_i^*(t))$.
Combining the two inequalities yields $\Gamma_e(\sigma_i'(t), \bm{x}_i'(t)) \leq (4+\epsilon) \Gamma_e(\sigma_i^*(t), \bm{x}_i^*(t))$.
Hence, we have obtained a solution whose execution cost is no larger than the $4+\epsilon$ times of the optimal cost.

Although the above solution provides an upper bound for the resulted execution cost, it is designed to optimize the worst-case performance,
which may not produce satisfying results in general cases.
To obtain a better average performance, we leverage the DRL methods to find the optimal processing order 
based on task information (i.e. workload and punishment coefficient of latency) and 
environmental information (i.e. price and available amount of computational resources).
Our approach is based on the DRL algorithm \cite{bello2016neural} designed to solve the Traveling Salesman Problem (TSP) problem,
where one needs to find an optimal sequence of cities with minimal tour length.
Notice that the objective of our problem is also computing an optimal ordering based on the attributes of tasks,
thus the neural network structure presented in \cite{bello2016neural} can be easily adapted to our situation.
The main difference between our problem and TSP comes from two aspects:
(i) in addition to the attributes of tasks, the optimal solution of our problem also depends on the environmental information;
(ii) the output ordering must be fed into the dynamic programming module to obtain the final execution cost.
These two problems can be solved by (i) combining task information with environmental information at the input layer
and (ii) connecting the dynamic programming module with the DRL module during the training process.

\begin{figure*}[!t]
\centering
\subfloat[First Stage]{\includegraphics[width=0.32\textwidth]{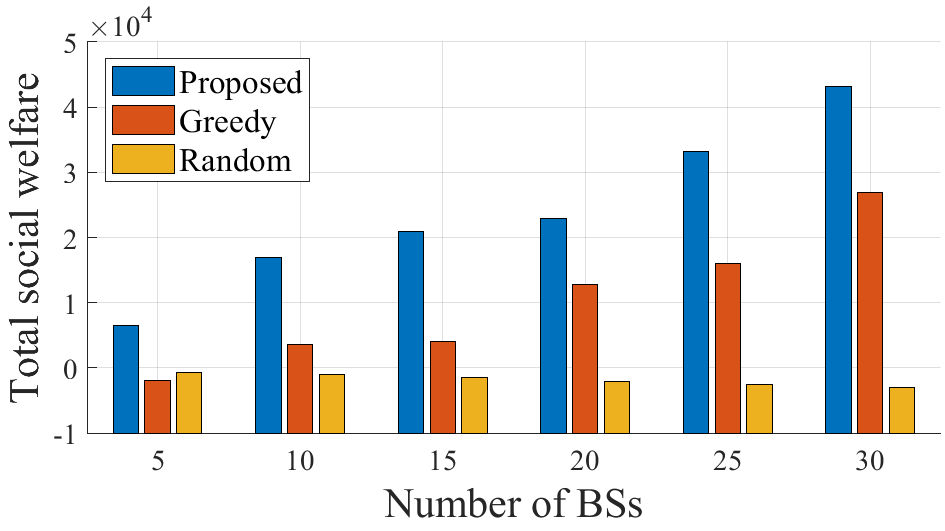} \label{fig:stage1}}
\hfil
\subfloat[Second Stage]{\includegraphics[width=0.32\textwidth]{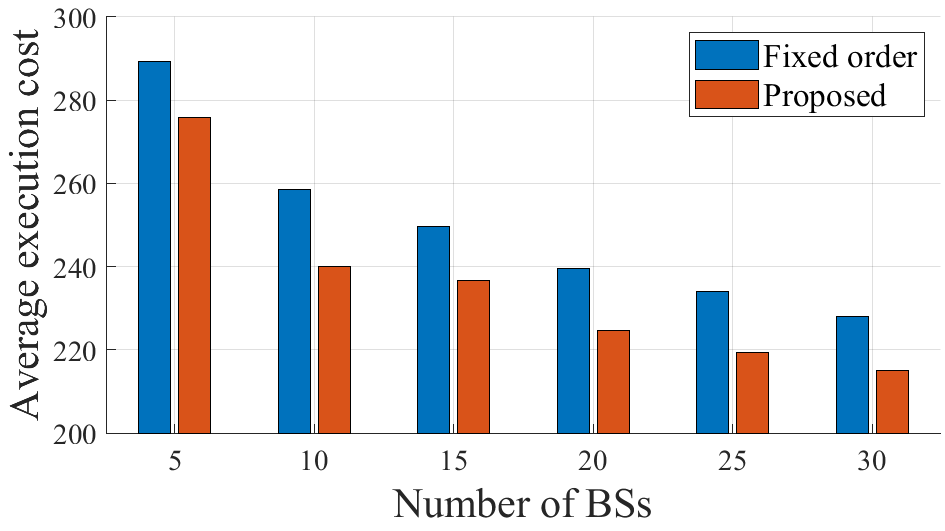} \label{fig:stage2}}
\hfil
\subfloat[Global]{\includegraphics[width=0.32\textwidth]{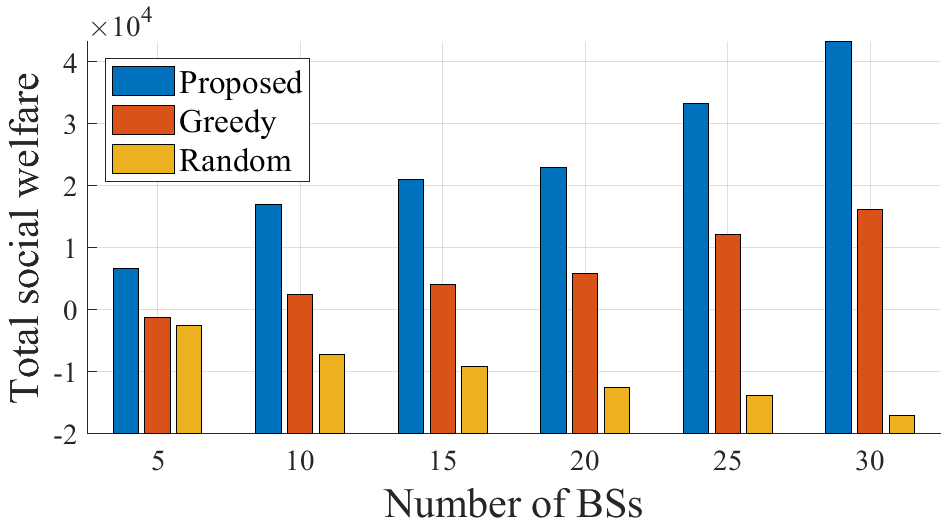} \label{fig:global}}
\caption{Performance under different numbers of BSs.}
\label{fig:sw}
\end{figure*}

\section{Numerical Results} \label{section:simulation}
In this section, extensive simulations are conducted to evaluate the performance of the proposed algorithm.
We consider a CEC network consisting of $N=10$ BSs and each experiment lasts for $200$ slots.
The computational power of each BS ranges from $20$GHz to $40$GHz and the workload on each BS ranges from $50\%$ to $120\%$.
When the BS's workload exceeds $100\%$, it posts offloading requests until the workload decreases to $100\%$.
When the BS's workload is less than $80\%$, it shares its spare computational resources for extra profits.
The price of resources is proportional to the reciprocal of the resources' available amount.
At every time slot, the BSs report the resource information over the next $\delta=10$ slots.
The workload of each task $w_{\tau}$ is drawn randomly from $[5M, 20M]$.
The maximum utility $u^0_{\tau}$ and punishment coefficient $\alpha_{\tau}$ satisfy uniform distribution on
$[100, 500]$ and $[10, 90]$, respectively.
In the first stage of our algorithm, tasks are divided into groups containing $K=5$ tasks.
Our algorithm is compared with the following two benchmarks:
\begin{itemize}
    \item \emph{Greedy}:
        For each offloading request, we iterate all consecutive slots on each BS and choose the task scheduling with
        the largest surplus.
    \item \emph{Random}:
        For each offloading request, we first randomly choose a target BS.
        If the BS has sufficient spare resources, we randomly select a feasible task execution.
\end{itemize}

Fig. \ref{fig:sw} shows the algorithms' performance under different numbers of BSs.
To demonstrate the advantage of our algorithm in each stage, 
in Fig. \ref{fig:stage1}, the three algorithms only make decisions for the first stage 
and the second stage is determined by our second-stage algorithm proposed in Section \ref{subsection:second_stage}.
According to the results, the total social welfare produced by our algorithm and Greedy increases with the network scale.
The reason is twofold.
First, the number of offloading requests increases with the number of BSs, which leads to larger social welfare
if the surplus of each task is positive.
Second, with more BSs in the network, we have a higher chance to select cheaper resources for task scheduling.
However, since the Random algorithm usually makes unfavorable task allocation decisions, the average surplus of each task
is negative so it performs worse in larger networks.

In Fig. \ref{fig:stage2}, we compare our second-stage algorithm with the algorithm that adopts the
fixed processing order proposed in \cite{epstein2012universal}.
As discussed above, we can select cheaper resources when there are more BSs, so the average execution cost
decreases with the number of BSs.
However, our algorithm outperforms the other in all cases.
Fig. \ref{fig:global} presents the total social welfare achieved by the three algorithms.
Compared to Fig. \ref{fig:stage1}, the performance of Greedy and Random is degraded.
This shows that applying our second-stage algorithm to Greedy and Random can improve their performance,
which further demonstrates the effectiveness of our second-stage algorithm.

\begin{figure}[hb]
\centering
\subfloat[First Stage]{\includegraphics[width=0.23\textwidth]{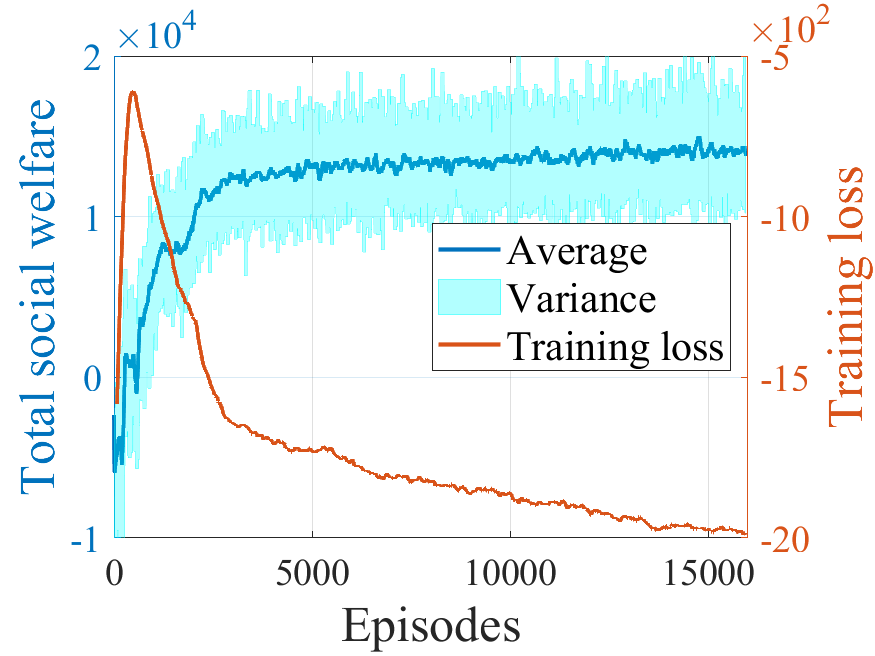} \label{fig:stage1_training}}
\hfil
\subfloat[Second Stage]{\includegraphics[width=0.23\textwidth]{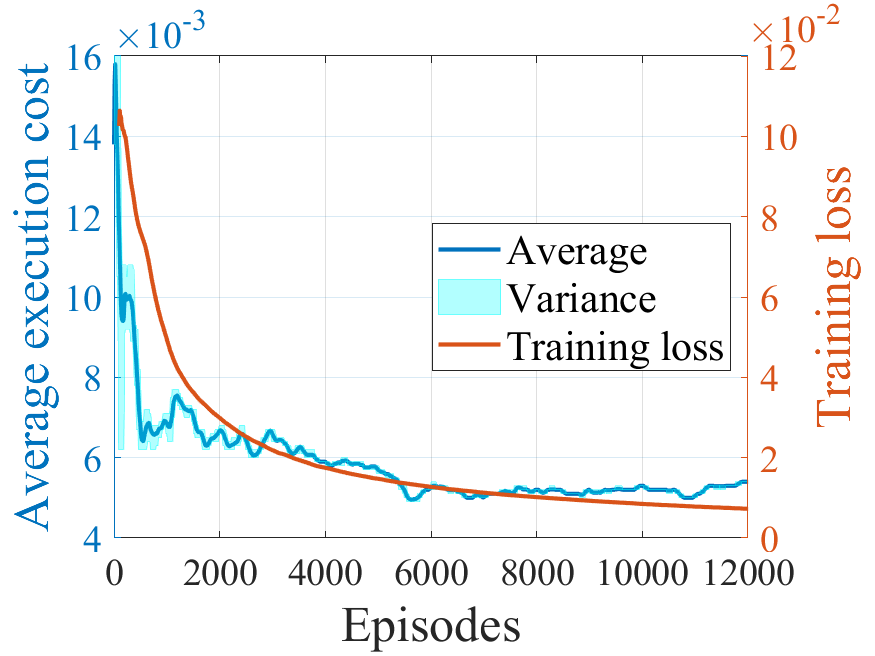} \label{fig:stage2_training}}
\caption{Algorithm performance and training loss under different training episodes.}
\label{fig:training}
\end{figure}

To show the convergence of our algorithm, we plot the algorithm performance and training loss for both stages in Fig. \ref{fig:training}.
The blue curve denotes the moving average of social welfare (or average cost) over the last $50$ episodes,
and the light blue shadow is the maximum and minimum value in the last $50$ episodes.
For the first stage, the social welfare and training loss change violently in the first $3000$ episodes and then gradually converge.
For the second stage, the average cost and training loss becomes stable after the $5000$-th episode.

\begin{table}[tb]
    \caption{Running time under different numbers of BSs.}
    \label{table:running_time}
\centering
\begin{tabular}{ccccccc}
\toprule
& \multicolumn{6}{c}{Number of BSs} \\
\cmidrule(lr){2-7}
Algorithm & N=5 & N=10 & N=15 & N=20 & N=25 & N=30 \\
\midrule
Proposed & 0.534 & 1.038 & 1.667 & 2.207 & 2.492 & 2.806 \\
Greedy & 0.062 & 0.235 & 0.447 & 0.862 & 1.075 & 1.713 \\
Random & 0.019 & 0.040 & 0.057 & 0.077 & 0.087 & 0.113 \\
\bottomrule
\end{tabular}
\end{table}

Table \ref{table:running_time} records the running time (in seconds) of each algorithm for $200$ slots under different numbers of BSs.
All experiments are conducted on a commodity personal computer and the running speed can be significantly improved
when utilizing dedicated hardware \cite{guo2004quantitative}.
Under a small network scale, the running time of our algorithm for a single slot is less than $3$ ms.
Notice that this value only grows linearly with respect to the number of BSs, 
hence our algorithm is applicable to large-scale networks.

\section{Conclusion} \label{section:conclusion}
In this paper, we have investigated cooperative mechanisms for CEC networks.
To maximize social welfare, an orchestrator has been designed to collect information and compute optimal task scheduling decisions.
We have formulated the scheduling problem as an MDP and proposed a two-stage DRL-based algorithm to solve it.
Numerical results have shown that our algorithm achieves a significant performance improvement compared with the benchmark schemes.
For the future work, we will combine our mechanism with the blockchain technology to improve its reliability and security.

\appendices

\section{Proof of Theorem \ref{theorem:useup}}
\label{appendix:proof_useup}
According to Theorem \ref{theorem:no_simul}, we can assume tasks are processed one by one in $A^{e,*}_i(t)$.
For any task $\tau\in\mathcal{T}_i(t)$, we suppose that its starting and ending time slot are $t_s$ and $t_e$, respectively.
We first prove that there is an optimal task execution decision with at most one partially utilized slot in $[t_s, t_e]$.
We prove by contradiction.
Suppose there are two partially used slots $t_1,t_2\in [t_s, t_e]$, i.e. $0 < \sum_{\tau\in\mathcal{T}_i(t)} f_{\tau}(t') < C^t_i(t')$ for $t'=t_1$ and $t'=t_2$.
If $p^t_i(t_1) \leq p^t_i(t_2)$, we reduce $f_{\tau}(t_2)$ and increase $f_{\tau}(t_1)$ by the same amount untill $f_{\tau}(t_2) = 0$
or $\sum_{\tau\in\mathcal{T}_i(t)} f_{\tau}(t_1) = C^t_i(t_1)$.
Similarly, if $p^t_i(t_1) > p^t_i(t_2)$, we adjust $f_{\tau}$ in the contrary way.
Notice that such manipulation reduces the execution cost of $\tau$ and does not increase its latency.
Hence, we have resulted in a new task execution decision $A^{e,\prime}_i(t)$ where the surplus $r_{\tau}$ is increased and the execution 
of other tasks remains unchanged.
As a result, we have $R^e(A^{e,\prime}_i(t)) \geq R^e(A^{e,*}_i(t))$.
Therefore, we either yield a contradiction or obtain another optimal task execution decision such that there is only one partially utilized slot in $[t_s, t_e]$.

Based on the above result, we can construct a new task execution decision $A^{e,\prime\prime}_i(t)$ by renting the remaining computation power
of all partially utilized slots in $A^{e,*}_i(t)$ to process some dummy tasks.
Since we have at most one partially utilized slot for each task, we have 
\begin{equation}
    R^e(A^{e,*}_i(t)) - R^e(A^{e,\prime\prime}_i(t)) \leq |\mathcal{T}_i(t)| C^{t,max}_i p^{t,max}_i, \label{eq:appendix1}
\end{equation}
where $C^{t,max}_i = \max_{t'\in[t,t+\delta-1]}C^t_i(t')$ and $p^{t,max}_i = \max_{t'\in[t,t+\delta-1]}p^t_i(t')$.
Since $A^{e,\prime\prime}_i(t)$ is binary, we must have $R^e(\hat{A}^{e,*}_i(t)) \geq R^e(A^{e,\prime\prime}_i(t))$.
Substituting into \eqref{eq:appendix1} yields
\begin{equation}
    R^e(A^{e,*}_i(t)) - R^e(\hat{A}^{e,*}_i(t)) \leq |\mathcal{T}_i(t)| C^{t,max}_i p^{t,max}_i. \label{eq:appendix1_2}
\end{equation}
The processed workload in slot $t'$ is $\sum_{\tau\in\mathcal{T}_i(t)} f_{\tau}(t') \Delta t$.
Hence, if we reduce the duration of each slot, the unit price of frequency $p^t_i(t')$ should be reduced proportionally,
so we have $\lim_{\Delta t\to 0} p^{t,max}_i = 0$.
Substituting into \eqref{eq:appendix1_2} proves the theorem.

\bibliographystyle{IEEEtran}
\bibliography{ref}

\begin{thebibliography}{10}
\providecommand{\url}[1]{#1}
\csname url@samestyle\endcsname
\providecommand{\newblock}{\relax}
\providecommand{\bibinfo}[2]{#2}
\providecommand{\BIBentrySTDinterwordspacing}{\spaceskip=0pt\relax}
\providecommand{\BIBentryALTinterwordstretchfactor}{4}
\providecommand{\BIBentryALTinterwordspacing}{\spaceskip=\fontdimen2\font plus
\BIBentryALTinterwordstretchfactor\fontdimen3\font minus
  \fontdimen4\font\relax}
\providecommand{\BIBforeignlanguage}[2]{{%
\expandafter\ifx\csname l@#1\endcsname\relax
\typeout{** WARNING: IEEEtran.bst: No hyphenation pattern has been}%
\typeout{** loaded for the language `#1'. Using the pattern for}%
\typeout{** the default language instead.}%
\else
\language=\csname l@#1\endcsname
\fi
#2}}
\providecommand{\BIBdecl}{\relax}
\BIBdecl

\bibitem{hu2015mobile}
Y.~C. Hu, M.~Patel, D.~Sabella, N.~Sprecher, and V.~Young, ``Mobile edge
  computing—a key technology towards 5g,'' \emph{ETSI white paper}, vol.~11,
  no.~11, pp. 1--16, 2015.

\bibitem{tran2017collaborative}
T.~X. Tran, A.~Hajisami, P.~Pandey, and D.~Pompili, ``Collaborative mobile edge
  computing in 5g networks: New paradigms, scenarios, and challenges,''
  \emph{IEEE Communications Magazine}, vol.~55, no.~4, pp. 54--61, 2017.

\bibitem{shi2016edge}
W.~Shi, J.~Cao, Q.~Zhang, Y.~Li, and L.~Xu, ``Edge computing: Vision and
  challenges,'' \emph{IEEE internet of things journal}, vol.~3, no.~5, pp.
  637--646, 2016.

\bibitem{wang2018enabling}
K.~Wang, H.~Yin, W.~Quan, and G.~Min, ``Enabling collaborative edge computing
  for software defined vehicular networks,'' \emph{IEEE Network}, vol.~32,
  no.~5, pp. 112--117, 2018.

\bibitem{he2021shapley}
X.~He, X.~Wang, S.~Wang, S.~Xu, J.~Ren, C.~He, and Y.~Zhang, ``A shapley
  value-based incentive mechanism in collaborative edge computing,'' in
  \emph{2021 IEEE Global Communications Conference (GLOBECOM)}.\hskip 1em plus
  0.5em minus 0.4em\relax IEEE, 2021, pp. 1--7.

\bibitem{hou2021incentive}
W.~Hou, H.~Wen, N.~Zhang, J.~Wu, W.~Lei, and R.~Zhao, ``Incentive-driven task
  allocation for collaborative edge computing in industrial internet of
  things,'' \emph{IEEE Internet of Things Journal}, vol.~9, no.~1, pp.
  706--718, 2021.

\bibitem{auer2008near}
P.~Auer, T.~Jaksch, and R.~Ortner, ``Near-optimal regret bounds for
  reinforcement learning,'' \emph{Advances in neural information processing
  systems}, vol.~21, 2008.

\bibitem{ddpg}
\BIBentryALTinterwordspacing
T.~P. Lillicrap, J.~J. Hunt, A.~Pritzel, N.~Heess, T.~Erez, Y.~Tassa,
  D.~Silver, and D.~Wierstra, ``Continuous control with deep reinforcement
  learning.'' in \emph{ICLR (Poster)}, 2016. [Online]. Available:
  \url{http://arxiv.org/abs/1509.02971}
\BIBentrySTDinterwordspacing

\bibitem{watkins1992q}
C.~J. Watkins and P.~Dayan, ``Q-learning,'' \emph{Machine learning}, vol.~8,
  no. 3-4, pp. 279--292, 1992.

\bibitem{silver2014deterministic}
D.~Silver, G.~Lever, N.~Heess, T.~Degris, D.~Wierstra, and M.~Riedmiller,
  ``Deterministic policy gradient algorithms,'' in \emph{International
  conference on machine learning}.\hskip 1em plus 0.5em minus 0.4em\relax PMLR,
  2014, pp. 387--395.

\bibitem{wang2005preemptive}
G.~Wang, H.~Sun, and C.~Chu, ``Preemptive scheduling with availability
  constraints to minimize total weighted completion times,'' \emph{Annals of
  Operations Research}, vol. 133, no. 1-4, pp. 183--192, 2005.

\bibitem{hohn2015performance}
W.~H{\"o}hn and T.~Jacobs, ``On the performance of smith’s rule in
  single-machine scheduling with nonlinear cost,'' \emph{ACM Transactions on
  Algorithms (TALG)}, vol.~11, no.~4, pp. 1--30, 2015.

\bibitem{epstein2012universal}
L.~Epstein, A.~Levin, A.~Marchetti-Spaccamela, N.~Megow, J.~Mestre,
  M.~Skutella, and L.~Stougie, ``Universal sequencing on an unreliable
  machine,'' \emph{SIAM Journal on Computing}, vol.~41, no.~3, pp. 565--586,
  2012.

\bibitem{chen2021optimal}
L.~Chen, N.~Megow, R.~Rischke, L.~Stougie, and J.~Verschae, ``Optimal
  algorithms for scheduling under time-of-use tariffs,'' \emph{Annals of
  Operations Research}, pp. 1--23, 2021.

\bibitem{bello2016neural}
I.~Bello, H.~Pham, Q.~V. Le, M.~Norouzi, and S.~Bengio, ``Neural combinatorial
  optimization with reinforcement learning,'' \emph{arXiv preprint
  arXiv:1611.09940}, 2016.

\bibitem{guo2004quantitative}
Z.~Guo, W.~Najjar, F.~Vahid, and K.~Vissers, ``A quantitative analysis of the
  speedup factors of fpgas over processors,'' in \emph{Proceedings of the 2004
  ACM/SIGDA 12th international symposium on Field programmable gate arrays},
  2004, pp. 162--170.

\end{thebibliography}

\end{document}